 \documentclass[prl,reprint,a4paper]{revtex4-1}
 \usepackage{axodraw_mod}
 \usepackage{amssymb,amsmath}
 \usepackage{graphicx}
 \usepackage{epic}
 \usepackage{verbatim}
 \usepackage{nicefrac}

 \newcommand{\jouref}[4]{{#1}{\bf #2} (#3) #4}
 \newcommand{\ibid}[3]{{ibid.\ }{\bf #1} (#2) #3}

\begin{document}

 \title{Superconductivity driven by the screening of long-distance 
Coulomb interaction}

 \author{Kosuke Odagiri}

 \affiliation{Electronics and Photonics Research Institute,
 National Institute of Advanced Industrial Science and Technology, 
 Tsukuba Central 2,
 1--1--1 Umezono, Tsukuba, Ibaraki 305--8568, Japan}

 \date{\today}

 \begin{abstract}

  The pair-fluctuation contribution reduces the electrostatic screening 
length in superconductivity as compared to the normal state.
  When a conductor possesses a static background charge distribution, 
superconductivity arises even in the absence of an explicit pairing 
interaction, such that the Coulomb repulsion is reduced and the total 
energy is lowered.
  We demonstrate that the superconducting gap increases with increased 
background charge at first, after which the mixing of the Higgs and 
plasma modes suppresses superconductivity in the pseudogap phase.
  This indicates that the mechanism may be relevant to the cuprates and 
iron pnictides.
  When the background charge is identified with the incoherent component 
of optical conductivity in the cuprates, our results reproduce the 
shape, size and position of the superconducting dome with zero free 
parameters.
  A superconducting critical temperature of about $1000$\,K is possible 
in ion-doped conductors.
 \end{abstract}

 \maketitle

 \section{Introduction}

  In this letter, we propose and explore a novel mechanism of SC 
(`superconductivity' or `superconducting'), where SC is driven by 
charge-screening.

  Quite surprisingly, this mechanism seems able to explain the universal 
features of the unconventional superconductors such as the cuprates 
\cite{highTc_first,highTc} and the iron pnictides \cite{FeAs}, 
quantitatively without free parameters.
  By `universal features', we mean: 1.\ small amount of carrier- or 
charge-doping creates an SC `dome'; 2.\ the maximum $T_\text{c}$ (transition 
temperature) is correlated with electronic energy scales such as 
the characteristic temperature $T_0$ of spin fluctuation \cite{tct0} or 
more simply $W/(m_*/m)$ where $W$ is proportional to the bandwidth and 
$m_*$ is the effective mass at the Fermi level \cite{yanagisawa}; 3.\ 
there is an `underdoped' pseudogap state.

\begin{figure}[ht]
  \begin{picture}(120,50)(0,0)
   \Text(20,45)[t]{(a)}
   \Text(60,37)[c]{\small quasiparticle}
   \Photon(20,20)(50,20){3}{3.5}
   \ArrowArcn(60,20)(10,0,180)
   \ArrowArcn(60,20)(10,180,360)
   \Photon(70,20)(100,20){3}{3.5}
  \end{picture}
  \begin{picture}(120,50)(0,0)
   \Text(20,45)[t]{(b)}
   \Photon(20,20)(50,20){3}{3.5}
   \Text(67,32)[l]{$G$, $H$}
   \DashCArc(60,20)(10,0,360){5}
   \Photon(70,20)(100,20){3}{3.5}
  \end{picture}

  \begin{picture}(120,50)(0,0)
   \Text(20,45)[t]{(c)}
   \Photon(20,10)(60,10){3}{4}
   \Photon(60,10)(100,10){-3}{4}
   \Text(72,20)[l]{$G$, $H$}
   \DashCArc(60,20)(10,0,360){5}
  \end{picture}
  \begin{picture}(120,50)(0,0)
   \Text(20,45)[t]{(d)}
   \Photon(20,20)(50,20){3}{3.5}
   \Text(67,32)[l]{$H$}
   \DashCArc(60,20)(10,0,180){5}
   \PhotonArc(60,20)(10,180,360){1.5}{3.5}
   \Photon(70,20)(100,20){3}{3.5}
  \end{picture}
  \caption{\label{fig_photon_self_energy}
  The diagrams for the self-energy of the photon (curly lines). Diagram 
(d) is zero by choice of gauge. Photonic tadpole graph, which is omitted 
here, is also zero by the choice of gauge}
 \end{figure}
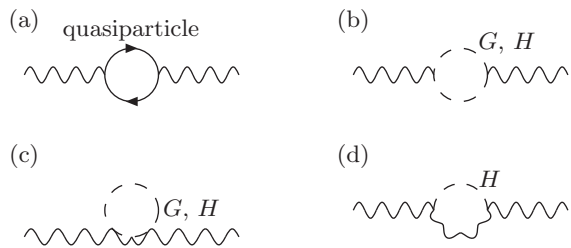

  Our mechanism is based on the reduction, typically by $\sim10\%$, of 
the electrostatic screening length in the SC phase. This is caused by 
pair fluctuation, which is calculated as loops of Higgs and Goldstone 
bosons, shown in Fig.~\ref{fig_photon_self_energy}.
  This effect is not surprising, because the SC condensate contributes 
extra degrees of freedom to charge screening compared to the normal 
state.

  A conductor that contains an unbalanced and static background charge 
distribution $\rho_\text{s}$ will therefore become SC such that the net 
Coulomb repulsion is reduced.
  The charge distribution $\rho_\text{s}$ may be either ionic, or due 
to some charge carrier that is immobilized by, e.g., spin or orbital 
interaction.

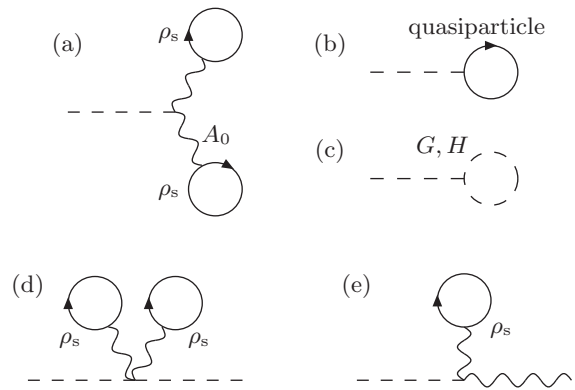
\begin{figure}[ht]
  \begin{center}
  \begin{picture}(105,80)(0,0)
   \Text(20,70)[t]{(a)}
   \Text(62,69)[r]{$\rho_\text{s}$}
   \Text(62,10)[r]{$\rho_\text{s}$}
   \Text(70,28)[bl]{$A_0$}
   \DashLine(20,40)(60,40){5}
   \Photon(60,40)(70,60){2.5}{2}
   \Photon(60,40)(70,20){2.5}{2}
   \ArrowArcn(75,69)(10,360,0)
   \ArrowArcn(75,11)(10,240,-120)
  \end{picture}
  \begin{picture}(105,80)(0,0)
   \Text(10,70)[t]{(b)}
   \DashLine(25,55)(60,55){5}
   \ArrowArcn(70,55)(10,270,-90)
   \Text(65,76)[t]{quasiparticle}
   \Text(10,30)[t]{(c)}
   \DashLine(25,15)(60,15){5}
   \DashCArc(70,15)(10,0,360){5}
   \Text(62,27)[r]{$G,H$}
  \end{picture}

  \begin{picture}(120,60)(0,10)
   \Text(20,50)[t]{(d)}
   \DashLine(20,10)(60,10){5}
   \DashLine(60,10)(100,10){5}
   \Photon(60,10)(70,30){2.5}{2}
   \Photon(60,10)(50,30){2.5}{2}
   \ArrowArcn(75,39)(10,360,0)
   \ArrowArcn(45,39)(10,360,0)
   \Text(40,25)[r]{$\rho_\text{s}$}
   \Text(80,25)[l]{$\rho_\text{s}$}
  \end{picture}
  \begin{picture}(120,60)(0,10)
   \Text(20,50)[t]{(e)}
   \DashLine(20,10)(60,10){5}
   \Photon(60,10)(100,10){2.5}{3.5}
   \Photon(60,10)(60,30){2.5}{2}
   \ArrowArcn(60,40)(10,360,0)
   \Text(70,29)[l]{$\rho_\text{s}$}
  \end{picture}
  \end{center}
  \caption{\label{fig_slug}
  The `slug' (a), the quasiparticle (b) and bosonic (c) contributions to 
the Higgs tadpole; new contribution to the Higgs self-energy (d); and 
the Higgs--photon (plasmon) mixing diagram (e). $\rho_\text{s}$ refers 
to the background charge}
 \end{figure}

  Diagrammatically speaking, the energy gain due to this effect is 
described by Fig.~\ref{fig_slug}a. This needs to be balanced by the 
more conventional contributions of Figs.~\ref{fig_slug}b, c.
  There is also a self-energy contribution of Fig.~\ref{fig_slug}d.

 The gap $\Delta_\text{SC}$ increases with increased $\rho_\text{s}$ at 
first, after which the contribution of Fig.~\ref{fig_slug}e causes 
Higgs--plasma mixing and suppresses SC.
  The latter phase corresponds to the pseudogap, where SC pairs are 
formed for a short time but are trapped and dissolved by the background 
charge.

  Our calculation is based on nonperturbative (quasi-perturbative) 
one-loop self-consistency conditions of the long-distance Higgs and 
Goldstone modes, and has essentially zero free parameters.

  The results of our calculation in the two-dimensional (2-d) case are 
consistent with the experimental findings, provided that the 
`incoherent' component \cite{cuprates_optical} of optical conductivity 
acts as $\rho_\text{s}$.
  The size, position and shape of the SC dome are reproduced.

  For the three-dimensional (3-d) case, because of the smaller size of 
SC fluctuation, the SC gap will be nearly three times greater than the 
2-d case and will typically be $\mathcal{O}$(0.1\,eV).
  The fabrication of a bulk material with charge imbalance will be a 
technological challenge.
  For proof-of-principle studies, a more practical alternative will be 
to study surface conductivity under applied electric field 
\cite{electric_field_induced}.

 \section{Definitions and calculational setup}
 \label{sec_definitions}

  The SC gap is defined by:
 \begin{equation}
  \mathcal{L}=
  -\Psi^\dagger
  \left(\kern-0.4em\begin{array}{cc} \xi(\mathbf{k}) &
  e^{-i\phi_\text{SC}}\Delta_\text{SC} \\
  e^{+i\phi_\text{SC}}\Delta_\text{SC} &
  -\xi(-\mathbf{k}) \end{array} \kern-0.4em\right) \Psi.
  \label{eqn_superconducting_dispersion_relation}
 \end{equation}
  We adopt $\phi_\text{SC}=0$ without loss of generality.
  $\Psi$ is the Nambu doublet \cite{namburepresentation}, and is 
parametrized in terms of the lower and upper quasiparticle bands as:
 \begin{equation}
  \Psi=
  \left(\kern-0.4em\begin{array}c \psi_\uparrow \\ \psi^*_\downarrow
  \end{array}\kern-0.4em\right)=
  \left(\kern-0.4em\begin{array}{cc}
  \cos\theta_\text{SC} & -\sin\theta_\text{SC} \\
  \sin\theta_\text{SC} & \cos\theta_\text{SC}
  \end{array}\kern-0.4em\right)
  \left(\kern-0.4em\begin{array}{c}u\\\ell\end{array}\kern-0.4em\right).
 \end{equation}

  Equation~(\ref{eqn_superconducting_dispersion_relation}) is 
diagonalized by mixing angle $\theta_\text{SC}$ given by
 \begin{equation}
  \tan2\theta_\text{SC}(\mathbf{k})=
  \Delta_\text{SC}/\xi(\mathbf{k}),
  \quad (-\pi/4\le\theta_\text{SC}\le+\pi/4).
 \end{equation}

  We generalize eqn.~(\ref{eqn_superconducting_dispersion_relation}) by 
including the Higgs and Goldstone degrees of freedom. We introduce the 
following parametrization:
 \begin{equation}
  \Phi_\text{SC}=(v+H)\sigma_1+G\sigma_2.
 \end{equation}
  $v$ is the Higgs-boson vacuum expectation value. $\sigma$ are the 
Pauli matrices. $G$ corresponds to the broken part of the vectorial U(1) 
symmetry under
 \begin{equation}
  \Psi\to e^{i\theta_\text{V}\sigma_3} \Psi,
 \end{equation}
  whereas $H$ corresponds to the conserved axial U(1)$_\text{A}$ 
symmetry under
 \begin{equation}
  \Psi\to e^{i\theta_\text{A}} \Psi.
 \end{equation}

  The interaction of the Higgs and Goldstone bosons with fermions is 
given by the following generalization of the off-diagonal part of 
eqn.~(\ref{eqn_superconducting_dispersion_relation}):
 \begin{equation}
  \mathcal{L}_{\Phi\psi\psi}=
  -v^{-1}\Delta_\text{SC} \Psi^\dagger\Phi_\text{SC}\Psi.
  \label{eqn_psipsiphi}
 \end{equation}

  $H$ and $G$ kinetic terms are parametrized by
 \begin{eqnarray}
  \mathcal{L}_\text{kin}&=&
  \frac14\,\text{Tr}\left|\left(i\hbar\frac{\partial}{\partial t}+
  2eA_0\sigma_3\right)\Phi_\text{SC}\right|^2
  \nonumber\\&-&
  \frac{u^2}{4}\,\text{Tr}\left|\left(-i\hbar\nabla+
  2e\mathbf{A}\sigma_3\right)\Phi_\text{SC}\right|^2,
  \label{eqn_kinetic_term}
 \end{eqnarray}
  $e$ is positive in our convention.
  This is similar to the standard Ginzburg--Landau formulation where the 
one-component order-parameter $\Phi_\text{GL}$ is proportional to 
$(v+H)+iG$, and has the advantage that the unphysical $H$--$A_0$ mixing 
term is absent.

  After the symmetry breaking, the electrostatic plasma mode develops an 
energy gap $\Delta_{A_0}=2ev$, whereas the magnetic mode acquires an 
energy gap $\Delta_\mathbf{A}=2(u/c)ev$, which corresponds to the 
penetration depth $\lambda_\text{SC}=c/\Delta_\mathbf{A}=c^2/2uev$.

  The symmetry-breaking potential is assumed to be of the form
 \begin{equation}
  V=
  (\Delta_H^2/8v^2)\left[G^2+(H+v)^2-v^2\right]^2.
  \label{eqn_multiboson_lagrangian}
 \end{equation}
  $\Delta_H$ is the excitation energy of the Higgs mode, and is related 
to the coherence length by
 \begin{equation}
  \xi_\text{SC}=u\hbar\sqrt2/\Delta_H.
  \label{eqn_coherence_length}
 \end{equation}

  Equation~(\ref{eqn_kinetic_term}) contains the following bilinear 
terms:
 \begin{equation}
  2ev\left[A_0\frac{\partial G}{\partial t}+
  u^2\mathbf{A}\cdot\nabla G\right].
  \label{eqn_bilinear_terms}
 \end{equation}
  We cancel these up to total derivatives by adding 't~Hooft 
gauge-fixing terms:
 \begin{equation}
  -\frac1{2\xi_g}\left(
  \frac{\partial A_0}{\partial t}+u^2\nabla\cdot\mathbf{A}
  -2ev\xi_g G\right)^2.
  \label{eqn_thooft_gauge}
 \end{equation}
  The subscript $g$ refers to gauge-fixing.
  For the situations concerning gauge-fixing in ordinary time-dependent 
Ginzburg--Landau theory, see, e.g., ref.~\cite{QDu}.

  The Goldstone and Higgs propagators are now given by
 \begin{eqnarray}
  D_G(E,\mathbf{k})&=&(E^2-u^2\mathbf{k}^2-\xi_g\Delta_{A_0}^2+i0)^{-1},
  \label{eqn_goldstone_propagator}\\
  D_H(E,\mathbf{k})&=&(E^2-u^2\mathbf{k}^2-\Delta_H^2+i0)^{-1}.
  \label{eqn_higgs_propagator}
 \end{eqnarray}

  The Goldstone-boson excitation energy $\xi_g^{1/2}\Delta{A_0}$ 
vanishes when $\xi_g\to0$. Since physical quantities should not depend 
on the gauge choice, let us choose this particular limit.
  This corresponds to assigning all longitudinal component of $A_\mu$ to 
$G$, where being longitudinal refers to being proportional to 
$(q_0,(u/c)^2\mathbf{q})\approx(q_0,\mathbf{0})$.
  In other words, $A_0$ is now described by $G$ whereas $\mathbf{A}$ is 
small for our purposes.
  This is with the exception of the case of $A_0$ coupling to a 
conserved electromagnetic current such as $\rho_\text{s}$, in which 
case the $A_0$ mode will propagate because the longitudinal coupling to 
$\rho_\text{s}$ gives zero.

  The Goldstone-boson contribution will yield logarithmic divergences in 
the 2-d case at finite temperatures in accordance with the 
Coleman--Mermin--Wagner theorem.
  We shall only study 2-d and 3-d systems at zero temperature here.

  The interaction Feynman rules are given by $i\mathcal{L}$, and are 
listed in Tab.~\ref{tab_feynman}.

 \begin{table}[ht] \begin{center}
 \begin{tabular}{cc}
 \toprule
 Vertex & Feynman rule \\
 \colrule
 $u_1^\dagger\ell_2G,\ \ell_1^\dagger u_2G$ &
 $\mp v^{-1}\Delta_\text{SC}\cos(\theta_1-\theta_2)$ \\
 $u_1^\dagger u_2G,\ \ell_1^\dagger\ell_2G$ &
 $\pm v^{-1}\Delta_\text{SC}\sin(\theta_1-\theta_2)$ \\
 $u_1^\dagger\ell_2 H,\ \ell_1^\dagger u_2H$ &
 $-iv^{-1}\Delta_\text{SC}\cos(\theta_1+\theta_2)$ \\
 $u_1^\dagger u_2H,\ \ell_1^\dagger\ell_2H$ &
 $\mp iv^{-1}\Delta_\text{SC}\sin(\theta_1+\theta_2)$ \\
 \colrule
 $u_1^\dagger\ell_2A_0,\ \ell_1^\dagger u_2A_0$ &
 $-ie\sin(\theta_1+\theta_2)$ \\
 $u_1^\dagger u_2A_0,\ \ell_1^\dagger\ell_2A_0$ &
 $\pm ie\cos(\theta_1+\theta_2)$ \\
 \colrule
 $GGH$ & $-iv^{-1}\Delta_H^2$ \\
 $HHH$ & $-3iv^{-1}\Delta_H^2$ \\
 $GGHH$ & $-iv^{-2}\Delta_H^2$ \\
 $GGGG,\ HHHH$ & $-3iv^{-2}\Delta_H^2$ \\
 \colrule
 $A_0A_0H$ & $+iv^{-1}\Delta_{A_0}^2$ \\
 $\mathbf{A}\mathbf{A}H$ & $-iv^{-1}\Delta_{\mathbf{A}}^2$\\
 $A_0A_0GG,\ A_0A_0HH$ & $+iv^{-2}\Delta_{A_0}^2$ \\
 $\mathbf{A}\mathbf{A}GG,\ \mathbf{A}\mathbf{A}HH$ &
 $-iv^{-2}\Delta_{\mathbf{A}}^2$\\
 \colrule
 $A_0GH$ & $v^{-1}\Delta_{A_0}E_G$ \\
 $\mathbf{A}GH$ & $-v^{-1}u\Delta_{\mathbf{A}}\mathbf{k}_G$ \\
 \botrule
 \end{tabular}
 \caption{\label{tab_feynman}Interaction Feynman rules}
 \end{center} \end{table}

 \section{Electrostatic charge screening}
 \label{sec_screening}

  Let us calculate $v$ by calculating the self-energy of $A_0$.
  The relevant Feynman diagrams are shown in 
Fig.~\ref{fig_photon_self_energy}.
  We are interested mainly in the case of zero external energy and 
momenta, which corresponds to calculating the long-distance screening 
length.

  The contribution of Fig.~\ref{fig_photon_self_energy}a to $v^2$ is 
given by
 \begin{equation}
  4v^2_{\operatorname{(a)}}=2i\int\frac{d^{d+1}k}{(2\pi)^{d+1}}
  (\sin2\theta_\text{SC}(\mathbf{k}))^2
  G_u(k)G_\ell(k)
 \end{equation}
  at zero external 4-momenta.
  Since the $u$ states are empty and $\ell$ states are occupied, we 
obtain
 \begin{equation}
  v^2_{\operatorname{(a)}}=\frac14
  \int\frac{d^d\mathbf{k}(\sin2\theta_\text{SC}(\mathbf{k}))^2}
  {(2\pi)^d\sqrt{\Delta_\text{SC}^2+\xi^2}}\approx\frac{g_F}4,
  \label{eqn_form_factor_initial}
 \end{equation}
  where $g_\text{F}$ is the normal-state density of states at the Fermi 
surface.
  The charge screening due to the quasi-particle loop is, as is well 
known, the same as the normal-state Fermi--Thomas screening.

  The contribution of Fig.~\ref{fig_photon_self_energy}b is given by
 \begin{eqnarray}
  v^2_{\operatorname{(b)}}&=&i\int\frac{d^{d+1}k}{(2\pi)^{d+1}}
  k_0^2D_G(k)D_H(k)\nonumber\\
  &=&\frac1{2\Delta_H^2}
  \int\frac{d^d\mathbf{k}}{(2\pi)^d}\left[
  \sqrt{u^2\mathbf{k}^2+\Delta_H^2}-u\left|\mathbf{k}\right|\right].
 \end{eqnarray}
  We shall discuss the momentum cutoff of the integrals in due course.
  The parameter $u$ is calculable by evaluating the self-energies of $G$ 
and $H$ at finite spatial momenta, with the result that $u\approx 
v_\text{F}$ where $v_\text{F}$ is the Fermi velocity.
  We shall omit the details here.

  The following approximation and parametrization are convenient:
 \begin{equation}
  v^2_{\operatorname{(b)}}\approx\frac18
  \int\frac{d^d\mathbf{k}}{(2\pi)^d}\left[
  \frac1{u\left|\mathbf{k}\right|}+
  \frac1{\sqrt{u^2\mathbf{k}^2+\Delta_H^2}} \right]
  = \frac{g_G+g_H}4.
 \end{equation}

  The third contribution is given by
 \begin{equation}
  v^2_{\operatorname{(c)}}=\frac{i}2\int\frac{d^{d+1}k}{(2\pi)^{d+1}}
  \left[D_G(k)+D_H(k)\right]=\frac{g_G+g_H}2.
 \end{equation}
  Hence
 \begin{equation}
  v^2=v^2_{\operatorname{(a)}}+
  v^2_{\operatorname{(b)}}+v^2_{\operatorname{(c)}}\approx
  (g_\text{F}+3g_G+3g_H)/4.
  \label{eqn_form_factor_final}
 \end{equation}

 \section{The mechanism of superconductivity}

  In conventional superconductivity \cite{BCS,AGD}, one representation 
of the gap equation is the following all-order equation:
 \begin{equation}
  \begin{picture}(170,20)(0,0)
   \Text(2,10)[l]{$\Delta_\text{SC}=\psi_\uparrow$}
   \ArrowLine(50,10)(70,10)
   \GCirc(80,10){10}{0.7}
   \Text(80,10)[c]{1PI}
   \ArrowLine(110,10)(90,10)
   \Text(115,10)[l]{$\psi^*_\downarrow+\Delta_\text{SC}^\text{bare}$.}
  \end{picture}
  \label{eqn_bcs_self_energy}
 \end{equation}
  The blob contains all one-particle irreducible contributions.
  Since $\Delta_H$ is generated dynamically, the all-order versions of 
Figs.~\ref{fig_slug}b, c cancel automatically provided 
$\Delta_\text{SC}^\text{bare}\to0$.
  This justifies the conventional prescriptions.

  Let us consider the case where there is no direct pairing interaction 
as such, and SC is stabilized by Fig.~\ref{fig_slug}a.
  We then obtain $\Delta_\text{SC}=\Delta_\text{SC}^\text{bare}$, and 
$\Delta_\text{SC}$ is now determined by the cancellation of 
Figs.~\ref{fig_slug}a, c, which minimizes the ground-state energy.
  The contribution of Fig.~\ref{fig_slug}b is zero to the leading order.

  Let us carry out this programme.
  Figure~\ref{fig_slug}a gives the energy gain due to screening:
 \begin{equation}
  \mathcal{A}_{\operatorname{(a)}}=
  -\frac12\,v^{-1}\Delta_{A_0}^2\left(\frac1{\Delta_{A_0}^2}\right)^2
  \left(\rho_\text{s}e\right)^2=-\frac{v^{-3}\rho_\text{s}^2}8.
 \end{equation}
  Figure~\ref{fig_slug}c gives the energy loss due to bosonic Casimir 
energy or, in other words, SC fluctuation:
 \begin{eqnarray}
  \mathcal{A}_{\operatorname{(c)}}&=&
  -\frac12\int\frac{d^{d+1}k}{(2\pi)^{d+1}i}v^{-1}\Delta_H^2
  \left(D_G(k)+3D_H(k)\right)\nonumber\\
  &=&v^{-1}\Delta_H^2(g_G+3g_H)/2.
  \label{eqn_tadpole_bosonic}
 \end{eqnarray}
  Therefore the equilibrium condition is
 \begin{equation}
  \frac{\rho_\text{s}^2}{g_\text{F}+3g_G+3g_H}=
  (g_G+3g_H)\Delta_H^2.
  \label{eqn_slug_equilibrium}
 \end{equation}

  $H$ and $A_0$ states mix through the process shown in 
Fig.~\ref{fig_slug}c.
  The contribution equals $ev^{-1}\rho_\text{s}$, and so the 
$H$--$A_0$ mass matrix is given by
 \begin{equation}
  \frac12\left(H \; A_0\right)\left(\kern-0.4em\begin{array}{cc}
  \Delta_H^2 & ev^{-1}\rho_\text{s} \\
  ev^{-1}\rho_\text{s} & \Delta_{A_0}^2
  \end{array}\kern-0.4em\right)\left(
  \kern-0.4em\begin{array}{c}H\\A_0\end{array}\kern-0.4em\right).
 \end{equation}
  We redefined $A_0$ as positive norm here for the sake of convenience.

  After diagonalization, the rotated $\Delta_H^2$ is given by
 \begin{equation}
  \Delta_{H^\prime}^2=\Delta_H^2-v^{-4}\rho_\text{s}^2/4.
  \label{eqn_rotated_delh}
 \end{equation}
  We used $\Delta_{A_0}^2\gg\Delta_H^2$.
  Hence
 \begin{equation}
  \Delta_H^2=\frac{v^{-2}\rho_\text{s}^2}{4g_G+12g_H},\quad
  \Delta_{H^\prime}^2=\frac{v^{-4}\rho_\text{s}^2
  (v^2-g_G-3g_H)}{4g_G+12g_H}.
  \label{eqn_screening_sc_delh}
 \end{equation}

\begin{figure}[ht]
  \begin{picture}(120,45)(0,0)
   \Text(20,42)[t]{(a)}
   \Text(60,37)[c]{\small quasiparticle}
   \DashLine(20,20)(50,20){5}
   \ArrowArcn(60,20)(10,0,180)
   \ArrowArcn(60,20)(10,180,360)
   \DashLine(70,20)(100,20){5}
  \end{picture}
  \begin{picture}(120,45)(0,0)
   \Text(20,42)[t]{(b)}
   \Text(70,30)[l]{$G$, $H$}
   \DashLine(20,20)(50,20){5}
   \DashCArc(60,20)(10,0,360){5}
   \DashLine(70,20)(100,20){5}
  \end{picture}

  \begin{picture}(120,35)(0,0)
   \Text(20,32)[t]{(c)}
   \DashLine(20,5)(100,5){5}
   \Text(73,20)[l]{$G$, $H$}
   \DashCArc(60,15)(10,0,360){5}
  \end{picture}
  \caption{\label{fig_higgs_self_energy}
  The diagrams for the self-energy of the Higgs (and Goldstone) boson
  }
 \end{figure}
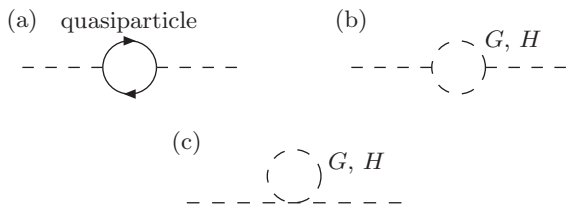

  To proceed further, we need to determine the cutoff of the integrals 
in $g_G$, $g_H$.
  This can be done by evaluating $\Delta_H$ at finite momenta. 
  In general, $\Delta_H$ is calculated using self-energy diagrams of 
Fig.~\ref{fig_higgs_self_energy} and Fig.~\ref{fig_slug}d.
  The contributions are individually divergent, but the sum is finite 
(except the long-distance and imaginary divergence of 
Fig.~\ref{fig_higgs_self_energy}b) so long as the tadpoles of 
Figs.~\ref{fig_slug}a, b, c cancel.

\begin{figure}[ht]
  \input{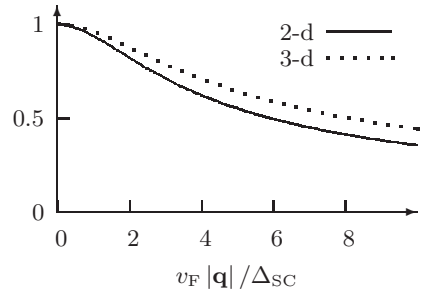}
  \caption{\label{fig_delh_q}
  $\Delta_H^2(0,\mathbf{q})/\Delta_H^2(0)$ as a function of 
$v_\text{F}\left|\mathbf{q}\right|/\Delta_\text{SC}$}
 \end{figure}

  After the cancellations take place, and at zero momentum, we obtain
 \begin{equation}
  \Delta_H^2(0)=g_\text{F}\int d\xi
  \frac{(v^{-1}\Delta_\text{SC}\sin2\theta_\text{SC})^2}
  {\sqrt{\Delta_\text{SC}^2+\xi^2}}=
  g_\text{F}\,v^{-2}\Delta_\text{SC}^2.
  \label{eqn_delh_zero}
 \end{equation}
  One way to generalize this to finite momenta is to calculate the 
difference of Higgs and Goldstone self-energies \cite{gribovewsb}:
 \begin{equation}
  \Delta_H^2(q)=\Sigma_H(q)-\Sigma_G(q).
 \end{equation}
  The dominant contribution is due to Fig.~\ref{fig_higgs_self_energy}a.
  This leads to a simple generalization of eqn.~(\ref{eqn_delh_zero}):
 \begin{equation}
  \Delta_H^2(0,\mathbf{q})=g_\text{F}\int d\xi
  \frac{(v^{-1}\Delta_\text{SC})^2
  \sin2\theta_\text{SC}\left<\sin2\theta^\prime_\text{SC}\right>}
  {\sqrt{\Delta_\text{SC}^2+\xi^2}}.
  \label{eqn_delh_q}
 \end{equation}
  Here $\tan2\theta_\text{SC}=\Delta/\xi(\mathbf{k})$ whereas
$\tan2\theta^\prime_\text{SC}=\Delta/\xi(\mathbf{k}+\mathbf{q})$.
  $\left<\dots\right>$ denotes the average over the solid angle of 
$\mathbf{k}$.
  For small $\mathbf{q}$ and an isotropic density of states, we obtain
 \begin{equation}
  \Delta_H^2(0,\mathbf{q})\propto\left<\text{Im}\left[
  \frac{\sin^{-1}\sqrt{z/2i}}{z\sqrt{1-2i/z}}\right]\right>,
 \end{equation}
  where $z=v_\text{F}\left|\mathbf{q}\right|\cos\phi/\Delta_\text{SC}$. 
  $\phi$ is the azimuthal angle of $\mathbf{q}$ with respect to 
$\mathbf{k}$.
  We integrate over the solid angle numerically, and obtain the results 
shown in Fig.~\ref{fig_delh_q}.
  The curves fall as $1/\left|\mathbf{q}\right|$ at large $\mathbf{q}$.
  We see that the half-width is approximately 6 for 2-d and 8 for 3-d.
  We shall use these numbers to estimate the cutoff.

 \section{Results}

  Let us now return to eqn.~(\ref{eqn_screening_sc_delh}).

  We introduce a simple cutoff at $u\left|\mathbf{k}\right|=c\Delta_H$ 
in $g_G$.
  Here $c\approx6$ for 2-d and $\approx8$ for 3-d.
  As $c$ is sufficiently large, $g_H=g_G$ to a good approximation.
  We obtain
 \begin{equation}
  g_G,\ g_H=
  \left\{\begin{array}{l}
   c\Delta_H/4\pi u^2 \quad (\text{2-d}), \vspace{2pt}\\
   c^2\Delta_H^2/8\pi^2 u^3 \quad (\text{3-d}).
  \end{array}\right.
 \end{equation}
  This implies
 \begin{equation}
  \Delta_H=
  \left\{\begin{array}{l}
   (u^2\rho_\text{s}^2\pi/4cv^2)^{1/3} \quad (\text{2-d}), 
   \vspace{2pt}\\
   (u^3\rho_\text{s}^2\pi^2/2c^2v^2)^{1/4} \quad (\text{3-d}).
  \end{array}\right.
 \end{equation}

  For SC to be stable, we need $\Delta_{H^\prime}^2>0$. This implies 
$g_\text{F}>g_G+9g_H$. Therefore, crudely speaking, $v^2\approx 
g_\text{F}/4$ by eqn.~(\ref{eqn_form_factor_final}), and 
$\Delta_H\approx2\Delta_\text{SC}$ by eqn.~(\ref{eqn_delh_zero}).
  There are thus two gaps, 
$2\Delta_\text{SC}\approx\Delta_H>\Delta_H^\prime$.
  It is natural to associate the former gap with the pseudogap.
  In this pseudogap phase, SC decays by coupling with $\rho_\text{s}$. 

\begin{figure}[ht]
 \centerline{\input{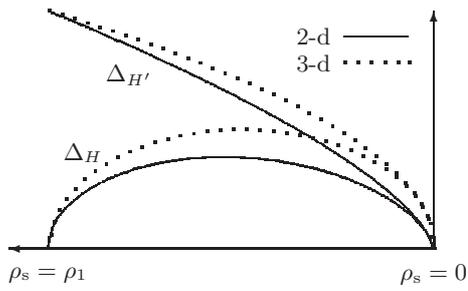}}
 \caption{\label{fig_universal_doping}
 $\Delta_H$ and $\Delta_{H^\prime}$ as function of $\rho_\text{s}$ 
(increases towards the left) in the 2-d and 3-d systems.
  With arbitrary normalization
 }
 \end{figure}

  Let us define $x_\text{s}=\rho_\text{s}/\rho_1$, where
 \begin{equation}
  \rho_1=
  \left\{\begin{array}{l}
   (2\pi/c)(uv^2)^2. \quad (\text{2-d}), 
   \vspace{2pt}\\
   (\pi/c)(2uv^2)^{3/2} \quad (\text{3-d}).
  \end{array}\right.
 \end{equation}
  We then obtain
 \begin{equation}
  \Delta_{H^\prime}=\frac{\rho_1}{2v^2}\times
  \left\{\begin{array}{l}
   x_\text{s}^{2/3} \sqrt{1-x_\text{s}^{2/3}} \quad (\text{2-d}),
   \vspace{2pt}\\
   \sqrt{x_\text{s}(1-x_\text{s})} \quad (\text{3-d}).
  \end{array}\right.
  \label{eqn_universal_dome}
 \end{equation}
  Adopting the approximation $v^2\approx g_\text{F}/4$, there is thus a 
universal overall behaviour for each number of dimensions as shown in 
Fig.~\ref{fig_universal_doping}.

  For the sake of estimation, let us adopt
 \begin{equation}
  (4v^2)u\approx g_\text{F} v_\text{F}\approx
  \left\{\begin{array}{l}
   k_\text{F}/\pi\approx1/a \quad (\text{2-d}), 
   \vspace{2pt}\\
   k_\text{F}^2/\pi^2\approx1/a^2 \quad (\text{3-d}).
  \end{array}\right.
 \end{equation}
  This yields
 \begin{equation}
  \rho_1\approx
  \left\{\begin{array}{l}
   0.065/a^2 \quad (\text{2-d}), 
   \vspace{2pt}\\
   0.14/a^3 \quad (\text{3-d}).
  \end{array}\right.
 \end{equation}
  We used $c=6$ for 2-d and $8$ for 3-d.
  From Fig.~\ref{fig_delh_q}, there is approximately $50\,\%$ 
uncertainty in the values of $c$ and hence in these numbers.

  Let us compare the results with the phenomenology in the 2-d case.
  For the cuprates, SC sets in at a doping of typically $0.05$ to $0.07$ 
per unit cell, and it has been found that at low doping, most of the 
doped carriers are `incoherent' \cite{cuprates_optical}.
  Supposing that these `incoherent' carriers are to be identified with 
$\rho_\text{s}$, these numbers agree well with $0.065/a^2$ derived 
above.
  SC in the cuprates disappears when the `incoherent' carriers vanish.
  This corresponds to $\rho_\text{s}=0$, and is again in agreement with 
the theory.

  As for the overall magnitude, let us define 
$uk_\text{F}\approx2W_\text{eff}$, the effective electronic energy scale 
as estimated at the Fermi level.
  We then obtain
 \begin{equation}
  \Delta_{H^\prime}^\text{max}\approx
  \left\{\begin{array}{l}
   0.032W_\text{eff} \quad (\text{2-d}), 
   \vspace{2pt}\\
   0.088W_\text{eff} \quad (\text{3-d}).
  \end{array}\right.
  \label{eqn_delhprime_max}
 \end{equation}
  For the cuprates, $W_\text{eff}\sim0.1$\,eV, and so the gap and hence 
$T_\text{c}$ is $\mathcal{O}$(50\,K), which agrees with the 
phenomenology.
  The universal correlation of $T_\text{c}$ with $W_\text{eff}$ is a 
well-known feature of a class of superconductors which includes the 
cuprates and iron pnictides.

 \section{Discussion}
 \label{sec_Discussion}

  The size of the coefficients in eqn.~(\ref{eqn_delhprime_max}) 
suggests that it will be surprisingly easy to achieve room-temperature 
SC.
  For $W_\text{eff}=1$\,eV for example, we would except a maximum $T_\text{c}$ 
of about 1000\,K in 3-d systems.

  The major factor which suppresses $W_\text{eff}$ and hence 
$T_\text{c}$ at present is the large size of the effective mass $m_*$.
  $m_*$ is typically large because the physics which creates the static 
trapped charge carriers $\rho_\text{s}$ also slows down the dynamical 
charge carriers.

  This situation needs not be the case in general, if our discussions 
have been correct.
  One can create $\rho_\text{s}$ more simply as impurity ions.

  For proof-of-principle studies, the simplest method would be to use an 
external electric field and study surface SC in 3-d bulk conductors that 
are doped with ions.
  The technology that is necessary for such a study will be similar to 
that adopted in the context of electric-field-induced SC 
\cite{electric_field_induced}.
  One would simply need to substitute an ion-doped conductor for 
SrTiO$_3$ that is studied in ref.~\cite{electric_field_induced}.
  The thickness of the surface layer will be governed by the screening 
length of the Coulomb interaction.

  As for the choice of the conductor, the only theoretical requirement 
is that the long-distance Coulomb interaction must not be suppressed.
  An isolated conduction band is therefore preferable, but $d$-band 
metals are permissible so long as the Fermi level is at the tail of the 
$d$-band, e.g., Cu or Zn.
  As for the dopant, ions that have the same sign of charge as the 
dynamical carriers would be preferable so as to avoid the trapping of 
dynamical carriers.
  For Cu, since the charge carriers are electrons, the dopant should be 
anions such as the halogens. For Zn, the dopant should be cations such 
as the alkali metals.
  The optimal doping concentration will be approximately 8\,$\%$.

 \section{Conclusions}
 \label{sec_conclusions}

  The suppression of long-distance Coulomb interaction in the SC phase 
implies that a conductor that is doped with static background charge 
will become SC in order that the total energy of the system is 
minimized.
  We studied this mechanism by utilizing a modified tadpole-cancellation 
condition.

  The SC gap increases with increasing background charge up to a certain 
point, after which SC is suppressed in the pseudogap phase when the 
mixing between the electrostatic plasmon field and the Higgs mode 
suppresses SC.

  The results are quantitatively consistent with the phenomenology for 
2-d systems.
  3-d systems will exhibit higher $T_\text{c}$ and will be candidates 
for room-temperature superconductivity.

  A further test of our mechanism will be the measurement of 
Higgs--plasmon mixing by optical emission, but this will be difficult 
because of the large size of the decay width 
$\Gamma_{H^\prime}\sim\Delta_{H^\prime}$.

  It would be interesting but challenging to amalgamate our 
long-distance formulation with a short-distance theory as, for instance, 
our theory cannot predict the pairing symmetry as it stands.
  This short-distance theory will be one that is equipped with a static 
charge distribution $\rho_\text{s}$ and mobile charge carriers, 
together with some implementation of long-distance Coulomb interaction.

 \begin{acknowledgments}

 \textbf{Acknowledgments:} The author thanks the members of the 
Superconducting Electronics Group, AIST, as well as N.~Yamada and 
K.~Yamaji for their patience and support.
  A precursory work was carried out in 2012 at Punjabi University, 
Patiala.
  We thank R.~C.~Verma and the supporting staff members of Punjabi 
University, Patiala, who made the visit possible.

 \end{acknowledgments}

 \end{document}